\documentclass[12pt]{iopart}

\usepackage[dvips]{graphicx}
\usepackage{array}
\usepackage{amssymb}
\usepackage{hhline}
\usepackage{longtable}
\usepackage{dcolumn}% Align table columns on decimal point
\usepackage{bm}% bold math
\usepackage{subfigure}
\usepackage[usenames]{color}
\begin{document}

\title[Prisoner's
dilemma in structured scale-free networks]{Prisoner's dilemma in
structured scale-free networks}

\author{Xing Li$^1$${}^,$$^2$, Yonghui Wu$^1$${}^,$$^2$, Zhihai Rong$^3$, Zhongzhi Zhang$^1$${}^,$$^2$, Shuigeng Zhou$^1$${}^,$$^2$}

\address{$^1$ School of Computer Science, Fudan
University, Shanghai 200433, China}
\address{$^2$ Shanghai Key Lab of
Intelligent Information Processing, Fudan University, Shanghai
200433, China}
\address{$^3$ Department of Automation, Donghua University, Shanghai 201620, China}
 \eads{\mailto{yhwu@fudan.edu.cn}, \mailto{zhangzz@fudan.edu.cn}, \mailto{sgzhou@fudan.edu.cn}}
\begin{abstract}
The conventional wisdom is that scale-free networks are prone to
cooperation spreading. In this paper we investigate the cooperative
behaviors on the structured scale-free network. On the contrary of
the conventional wisdom that scale-free networks are prone to
cooperation spreading, the evolution of cooperation is inhibited on
the structured scale-free network while performing the prisoner's
dilemma (PD) game. Firstly, we demonstrate that neither the
scale-free property nor the high clustering coefficient is
responsible for the inhibition of cooperation spreading on the
structured scale-free network. Then we provide one heuristic method
to argue that the lack of age correlations and its associated
`large-world' behavior in the structured scale-free network inhibit
the spread of cooperation. The findings may help enlighten further
studies on evolutionary dynamics of the PD game in scale-free
networks.

\end{abstract}


%87.23.Ge Dynamics of social systems
%87.23.Kg Dynamics of evolution
%89.75.Hc Networks and genealogical trees
%02.50.Le Decision theory and game theory
%02.10.Ox Combinatorics; graph theory
%Uncomment for PACS numbers title message
%\pacs{00.00, 20.00, 42.10}
% Keywords required only for MST, PB, PMB, PM, JOA, JOB?
%\vspace{2pc}
%\noindent{\it Keywords}: Article preparation, IOP journals
% Uncomment for Submitted to journal title message
%\submitto{\JPA}
% Comment out if separate title page not required
\maketitle
%%%%%%%%%%%%%%%%%%%%%%%%%%%%%%%%%%%%%%%%%%%%%%%%%%%%%%%%%%%%%%%%%%%%

\section{Introduction}

Cooperation is an essential factor for the evolution of species. For
decades, scientists have been embarking on the problem of
understanding the emergence of
cooperation~\cite{Axelrod81,nowak,killingback99,sg04,Nowak05,Ohtsuki06}.
In the field of this investigation, evolution game theory has become
a powerful tool~\cite{evolGame,evolTheory,gametheory}. One of the
most frequently used metaphors is the prisoner's dilemma (PD)
game~\cite{Axelrod84} that is modeled to simplify individual
interactions when selfish actions provide a short-term higher
payoff. In each round two players are involved in the game. The two
strategies for one player are to become a cooperator or a defector.
A cooperator is someone who pays some cost for another individual to
receive a benefit. A defector has no cost and does not deal out
benefits. Cost and benefit are measured in terms of fitness. The
combinations of the strategies will be of great difference. When
both of the players choose to cooperate or to defect, each of the
two will get benefit $R$ or $P$, respectively. When one player
choose to cooperate and the other choose to defect, the cooperator
will get $S$, while the defector will get $T$. In the PD game, the
order for the payoffs with different combinations of strategies is
$T > R> P > S$. So the best strategy is to defect regardless the
opponent's strategy and assuming that strategies are allowed to
spread within the population according to their
payoffs~\cite{evolGame}.

Recently, biological experiments have demonstrated that the
evolution dynamics is related to the topological structure of the
media, on top of which the evolution dynamics are
performed~\cite{Kerr02}. Earlier
work~\cite{nowak,sg04,szabo98,abramson01} mainly focused on the
two-dimensional lattice, which is modeled for the homogeneous
network. In these studies, a player is constrained to play solely
with its nearest neighbors. All the studies reported that unlike in
unstructured populations, cooperators and defectors can coexist in
the lattice indefinitely. However, the regular network is not
suitable for modeling real networks since a lot of empirical
studies~\cite{falout99,jeong01,newman01,pastor04} have uncovered
that the degree distribution $P(k)$ of many real-life systems
complies with a power-law form $P(k)\sim k^{-\lambda}$ with $2 <
\lambda \leqslant 3$. The scale-free property shows that everyone
plays a different role in the network, which describes a kind of
social diversity~\cite{santos08}. A famous model for generating
scale-free network is the Barab\'asi-Albert (BA)
model~\cite{random}.

The evolution of cooperation on scale-free networks has been
explored in~\cite{sf,Gomez,szabo2007,poncela07,Rong07,WangWenXu07},
which exhibited that cooperation is the dominating trait throughout
the entire range of parameters. In this paper, we focus on the
evolution of cooperation on a structured scale-free network, which
demonstrates negative age-correlations~\cite{hcsf}. Firstly, we
introduce the algorithm for constructing the network. Secondly, we
simulate the PD game on the network, and compare the results with
that of the BA network. In contrast to the results previously
obtained for the BA network, we find that the structured scale-free
network inhibits the spread of cooperation. Finally, we provide a
heuristic argument to explain why the spread of cooperation is
inhibited on the network considered, which is justified by extensive
simulations.

\section{Models}

Firstly let us introduce the celebrated BA model: Initially there
are $m_0$ connected nodes in the network. At each time step, one
node is added into the network and links with $m$ $(m< m_0)$
existing nodes. The probability that an existing node $i$ acquires a
new link is $P(k_i)=\frac{k_i}{\sum_jk_j} $, where $k_i$ is the
degree of node $i$. Repeat this step until getting the desired
network size. In the BA model, on one hand, the chance that an old
node receiving a new link is proportional to it age; this phenomenon
is called ``preferential attachment", also named
``age-correlations". On the other hand, the long-range connections
generated by the process decrease the distance between the vertices,
leading to a small-world phenomenon: the average path length
increases logarithmically with the network size~\cite{barabasi02}.

The aging of nodes is particularly interesting. Recently, Klemm and
Egu\'{i}luz found that for the scientific citation
network~\cite{newman01}, the age-correlation is negative: the mean
citation rate of papers decreases with the increase of their
age~\cite{hcsf}, that is to say, old nodes have less probability to
obtain links than those nodes just added into the network. To
describe this phenomenon, they put forward a new model called the
structured scale-free network (also called highly clustered
scale-free network). Their model is built as follows: Initially
there is a complete graph with $m$ active nodes. At successive time
step, one active node is added into the network connected to all $m$
active nodes, then one of $m+1$ active nodes is randomly chosen and
deactivated with probability $P(k_j)=\frac{s}{a+k_j} $ with $a > 0$
being a constant, where $s$ is a normalization factor defined by
$s^{-1}=\sum_i{\frac{1}{a+k_i}}$, in which $i$ belongs to the set of
active nodes. Repeat the process until the network size reaches the
needed size.

According to the above process, we know the age correlations are
greatly suppressed since every round only the active nodes, not all
existing nodes, have the chance to acquire new links. In result, the
linear topology of the structured scale-free network with local
highly connected clusters forms a long chain~\cite{topofHCSF}. The
power law exponent is $\lambda = 2+\frac{a}{m} $. In this paper, we
take $a = m$, and thus the structured scale-free network has the
same degree distribution with the BA model with $\lambda=3
$~\cite{random}. The clustering coefficient of BA network approaches
to zero as network size grows to infinite, while in the structured
scale-free network it is an asymptotic value $\frac 5 6$ for the
case $a = m$.

After introducing the two models, we will study the evolutionary PD
game on both  models in the following section.

%%%%%%%%%%%%%%%%%%%%%%%%%%%%%%%%%%%%%%%%%%%%%%%%%%%%%%%%%%%%%%%%
\section {Simulations}
In this section, we implement the finite population analogue of
replicator dynamics in the following PD simulations. Following
~\cite{nowak}, we make $T =b
> 1$, $R = 1$, and $P = S = 0$ in the PD game, where $b$ represents
the temptation to defect, being typically constrained to the
interval $1<b\leqslant2 $. In each round of evolution, one node $i$
plays the PD game with its directly connected neighbors,
accumulating the payoff as $P_i$. Whenever the strategy of the
player $i$ with $k_i$ neighbors is to be updated, a neighbor $j$ is
randomly selected from $i$'s neighborhood. If $P_j > P_i$, the
chosen neighbor $j$ spreads its strategy to the player $i$ with
probability $\frac{P_j - P_i} { b(\mbox{max}(k_i, k_j))}$,
otherwise, the player $i$ holds its strategy~\cite{sf}. Simulations
were carried out for a population with $N = 10{\ }000$ individuals.
Initially, cooperator and defector strategies were distributed
randomly among the players. Equilibrium frequencies of cooperators
and defectors were obtained by averaging over $1\ 000$ generations
after a transient time of $50\ 000$ generations. The frequency of
cooperators is a function of the parameter $b$ for the PD game. Each
data point corresponds to $100$ simulations that are $10$ runs for
$10$ different realizations of the same type of network specified by
the appropriate parameters (the population size $N$ and the average
connectivity $z$).

\begin{figure}
\begin{center}
\includegraphics[width=0.6\textwidth]{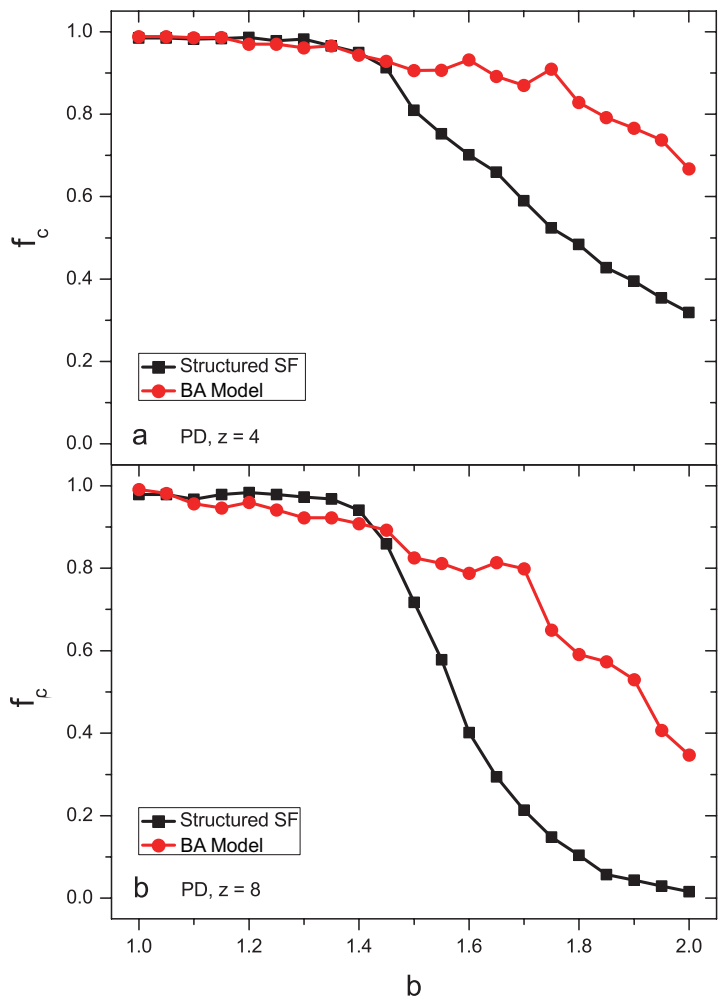} \
\end{center}
\caption[kurzform]{\label{Fig1} Frequency of cooperators $f_c$ as a
function of
   the advantage $b$ of defectors for the PD game. The lines with filled squares
   show the results of structured scale-free network in the PD game.
   The lines with dots are for the BA scale-free network. In the figure, $z$ is the average connectivity
   of the network. According the construction process of the structured scale-free network, we know $z = 2m$.
   The frequency of cooperation in the structured scale-free network
   is surpassed by that of the BA network in almost the entire range of $b$, especially when $b$ is large
   enough.
}
\end{figure}

Figure~\ref{Fig1} shows the results for the PD game on both networks
with different values of the average connectivity $z$. The BA
scale-free network and the structured scale-free network are
constructed through above methods. The frequency of cooperators,
$f_c$, is enhanced on the BA scale-free networks in the PD game,
dominating over the entire range of $b$ ($1\leqslant b \leqslant
2$). Notice that even when $2 < b \leqslant 3$, we have checked that
the results will not change. When $1<b\leqslant 1.4$, both of the
networks are all-cooperator networks. As the temptation value $b$
increases from $1.4$, the frequency of cooperators in the structured
scale-free network descends rapidly while the cooperative behavior
is always blooming in the BA network. Especially, for $b = 2$, when
$z = 4$, $f_c$ on the structured scale-free network is just half of
that on the BA network; when $z = 8$, $f_c$ decreases to $0$ on the
structured scale-free network, while the BA network still shows a
remarkable survival of cooperation with $f_c \approx 0.4$. On the
other hand, as $z$ grows, the advantage of scale-free network
shrinks. Therefore, it can be observed from figure \ref{Fig1} that
cooperators change into defectors more quickly with smaller
temptation value in the structured scale-free network, therefore, we
can conclude that comparing with the BA network, the structured
scale-free network decreases the frequency of cooperation, the
reason for which will be detailedly analyzed in the next section.

\section{Analysis}
\begin{figure}
\begin{center}
\includegraphics[width=0.6\textwidth]{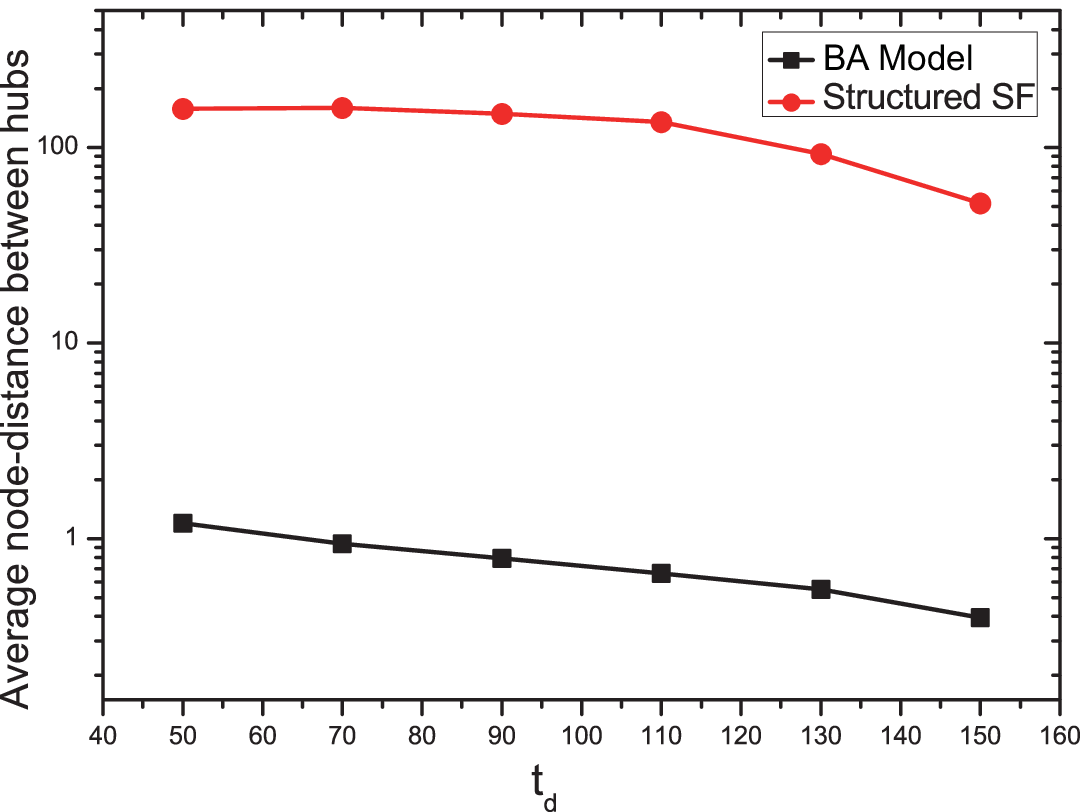} \
\end{center}
\caption[kurzform]{\label{Fig2} The average node-distance between
hubs as a function of the threshold of degree, $t_d$. A node $i$ is
called a hub node when its degree $k_i \geqslant t_d$. The solid
line with dot symbols is for the structured scale-free network and
the solid line with squares is for the BA network.}
\end{figure}

Why does the structured scale-free network inhibit the frequency of
cooperators comparing with BA network? Firstly, we are aware of that
both of the BA network and the structured scale-free network have
the same power law exponent ($\lambda = 3$) in the above
simulations. So the scale-free property is a trivial factor that can
not affect the simulation results. Secondly, the clustering
coefficient on the structured scale-free network is much higher than
that in BA network. Assenza et al. in ~\cite{Assenza08} find that
high clustering coefficient can help the enhancement of cooperation
on the scale-free networks with a tunable value of clustering
coefficient~\cite{Holme02}, which is a different network model with
the structured scale-free network. However, the high coefficient can
not help to enhance the cooperation spreading on the structured
scale-free network. So what is the root cause responsible for the
inhibition of cooperation spreading on the structured scale-free
network? From the models, we know the age correlations are greatly
suppressed on the structured scale-free network comparing with the
BA network, which results in the fact that, in the BA network, a hub
is usually linked to other highly connected nodes, while in the
structured scale-free network, a hub is almost exclusively connected
to low degree nodes~\cite{epicthreshold}. Also the diameter of the
structured scale-free network is increasing linearly with the
network size~\cite{topofHCSF}.

We can illustrate this phenomena through calculating the
node-distance for one hub to others in the scale-free network as
shown in figure~\ref{Fig2}. The value of node-distance from the
origin to the destination is defined as the number of nodes on the
shortest path between the two nodes. For example, the node-distance
of a pair of directly connected nodes is 0; if one of the shortest
paths between nodes $i$ and $j$ bypass another node $k$, then the
node-distance between $i$ and $j$ is $1$, and so on.
Figure~\ref{Fig2} shows the average node-distance between hubs as a
function of threshold $t_d$. A node $i$ can be qualified as a hub
node only if its degree $k_i \geqslant t_d$. We can see that in the
BA network, the hub nodes are usually interconnected with shorter
average node-distance. While in the structured scale-free network,
the average node-distance between hubs is rather high. When $t_d =
50$, the node-distance among hubs on the structured scale-free
network is more than $100$, while this value is just a slightly more
than one in the BA network. Hence, in the structured scale-free
network, the hubs are connected through some low degree nodes, which
form an ``intermediate region".

\begin{figure}
\begin{center}
\includegraphics[width=0.9\textwidth]{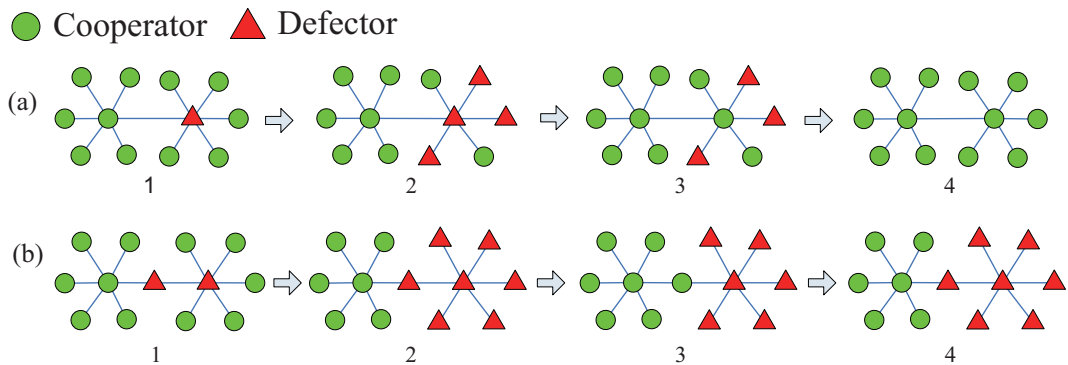} \
\end{center}
\caption[kurzform]{\label{Fig3} Evolution process on double-star
graphs. (a) Directly connected double-star for the BA network. The
centers of the two stars are directly connected. The left hub $h_l$
has $N - 1$ leaves and the right one  $h_r$ has $M - 1$ leaves. (b)
Double-star with intermediate region for the structured scale-free
network. Comparing with (a), the hubs are not directly connected,
but by the intermediate region. The degrees of hubs are the same
with (a). In this figure, we just put one node on the intermediate
region for the sake of simplifying analysis. }
\end{figure}

For the sake of simplicity, we model the two kinds of network
through two kinds of subgraph shown in figure~\ref{Fig3}, where (a)
shows a directly connected double-star for the BA scale-free network
and (b) is a double-star with intermediate region for the structured
scale-free network. In figure~\ref{Fig3}(a), without losing
generality, we assume that initially there are the unique defector
locating on the right hub $h_r$ and others(including the left hub
$h_l$) are cooperators. Set $T = b$, $R = 1$, and $P = S = 0$. In
the first generation of the networked PD game, the payoff of $h_l $
is $\pi(h_l) = N$ and the payoff of $h_r$ is $\pi(h_r) = Mb $.
According to the evolutionary rule, the probability for the defector
hub to invade the cooperator hub is

\begin{equation}\label{eq01}
P_1 = \frac{\pi(h_r) - \pi(h_l)}{b\mbox{max}(M, N)}= \frac{Mb -
N}{b\mbox{max}(M, N)}.
\end{equation}

Moreover, the payoff of some leaf on the right star $l_r $ is
$\pi(l_r) = 0 $ and the defector hub can also invade its leaves. The
probability for the defector hub to invade its leaf $l_r$ is

\begin{equation}\label{eq02}
P_2 = \frac{\pi(h_r) - \pi(l_r)}{b\mbox{max}(1, M)}= 1.
\end{equation}

Apparently, $P_1 < P_2 $, hence, the leaves aside the defector hub
are easier to be invaded by the defection than the cooperative hub.
After the defector hub has invaded some of its leaves and $k$ ($k <
M$) cooperator leaves survive, we get figure \ref{Fig3}(a-2) where
$\pi(h_r) = kb $. The probability that the defector hub invades the
cooperator hub changes into

\begin{eqnarray}\label{eq03}
P_3 &= \frac{\pi(h_r) - \pi(h_l)}{b\mbox{max}(M,
N)}=\frac{kb-N}{b\mbox{max}(M, N)}.
\end{eqnarray}

Comparing with $P_1$, it can be found that the cooperator hubs are
harder to be invaded by the defector hubs as the decrease of $k$.
When $k = 0$, i.e., all cooperator leaves aside the right star adopt
the defector strategy, $P_3$ reaches its minimal value and the
defector hub will definitely be invaded by the cooperator hub.
Hence, after $h_l$ changes into the cooperator, it will induce its
neighbors learning its behavior and the cooperative strategy will
spread to the whole double-star graph(figure~\ref{Fig3}(a-4)).

\begin{figure}
\begin{center}
\includegraphics[width=0.6\textwidth]{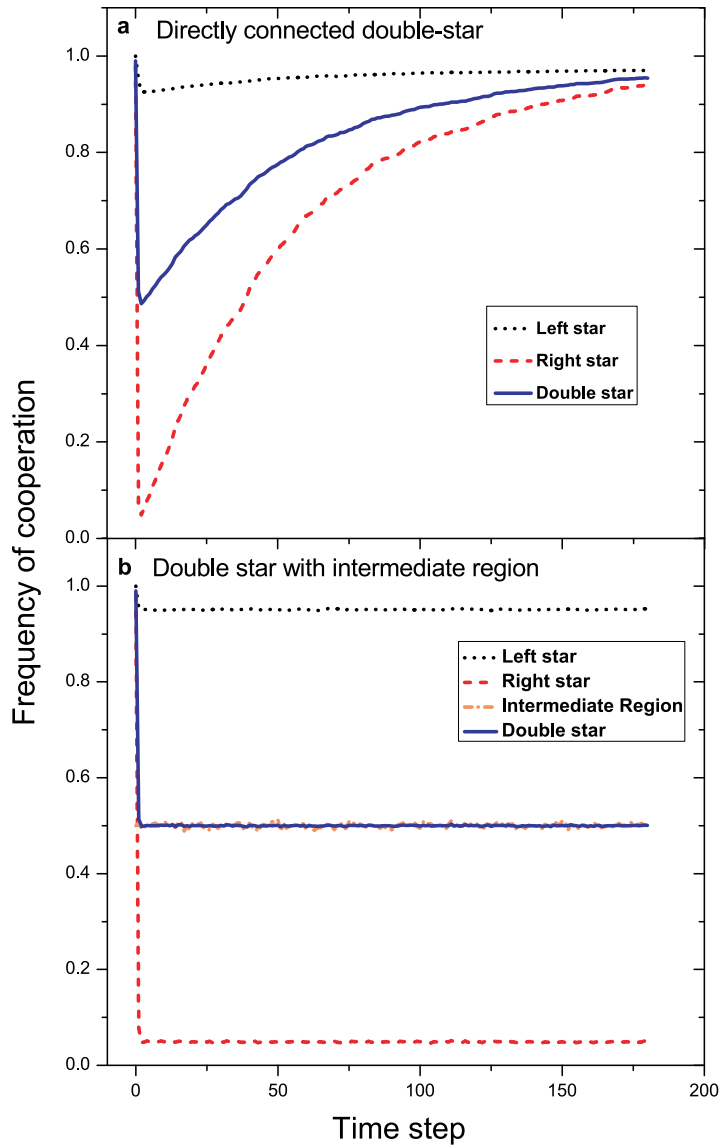} \
\end{center}
\caption[kurzform]{\label{Fig4} Evolution of cooperation under the
PD game on the double-stars graphs. Initially the cooperator hub
$h_l$ and the defector $h_r$  are linked with the same number of
leaves, $N = M = 50$. All leaves are cooperators. We fix $b = 1.5$.
The simulations are averaged over $1000$ runs. (a) The frequency of
cooperation $f_c$, on the left star (dot line) stays at nearly
$1.0$. $f_c$ in the right star (dashed line) first decreases to
nearly $0$ as defectors invades its leaves. Then at the lowest
point, $h_l$ spreads the cooperation strategy to $h_r$ successfully.
After that, the cooperators wipe up all the defectors on the right
star.  (b) We add a 10-node intermediate region between the
double-star as shown in the right corner. Initially the nodes on the
intermediate region randomly choose their strategies. The left star
contains all cooperators (dot line). The $f_c$ on the right star
(dash line) decreases to nearly $0$ (but not $0$ because of the
random effect) dramatically and stands still. The $f_c$ of
intermediate region (dash dot line) fluctuates around $0.5$ as
expected. The frequency of cooperation of the whole double star is
shown as solid line. }
\end{figure}

Through the structural analysis for the structured scale-free
network, we point out that the intermediate region plays the crucial
role for the evolution of cooperation. Hence, we add an intermediate
node among the two hubs as shown in figure~\ref{Fig3}(b) to
illustrate the effect of the intermediate region. We assume
initially the node x on the intermediate region is a defector and
others are the same as the figure \ref{Fig3}(a-1). At
figure~\ref{Fig3}(b-1), $\pi(h_l) = N $, $\pi(h_r) = 0 $ and $\pi(x)
= b $. If $x$ chooses $h_l$  to update its strategy, the probability
for $x$ to turn into a cooperator is

\begin{equation}\label{eq04}
P_4 = \frac{\pi(h_l) - \pi(x)}{b\mbox{max}(2, N)}= \frac{N-b}{bN}.
\end{equation}

Also according to the above analysis, the defector hub will invade
all its leaves, and then we get figure \ref{Fig3}(b-3). At this
point, $\pi(x) = 1 < \pi(h_r) = b $. It's impossible for $x$ to be
invaded by the defector hub with probability

\begin{equation}\label{eq05}
P_5 = \frac{\pi(h_r) - \pi(x)}{b\mbox{max}(2, M)}= \frac{b-1}{bM}.
\end{equation}

Therefore, the defector hub will receive the benefit from $x$ when
$x$ adopts the cooperative strategy, furthermore, it will invade $x$
and make $x$ become defector again. Then the process drops into a
loop shown between figure \ref{Fig3}(b-3) and figure
\ref{Fig3}(b-4), alternatively. The nodes on the intermediate region
fluctuate their strategies and the defector hub is protected by the
intermediate region against the invasion of cooperator hubs,
resulting in the fact that the cooperative behavior is inhabited on
the structured scale-free network for the PD game. In figure
\ref{Fig3}(b) we consider that the intermediate region only contains
one node, and in the case of multiple nodes on the intermediate
region the results do not change. Although we use the double-star
graphs to illustrate the cooperation evolution on the scale-free
networks, the analysis can heuristically reflect the microscopic
organization of cooperation on the evolutionary dynamic of the PD
game.

Furthermore, we have done extensive simulations shown in
figure~\ref{Fig4} to justify our above analysis. In the simulations
we bring in a kind of random effect~\cite{szabo2007}. We assume that
there are no direct connections among the leaves but adopt a
strategy from each other with probability $0.1$ which is the random
strategy adoption. From figure~\ref{Fig4} it is observed that for
the double-star graph, the defector hubs can be invaded by the
cooperator hub and after which, its neighbors also become
cooperators again. Whereas, for the double-star graph with
intermediate region, the leaves around the cooperative hub are
cooperators and that around the defector hubs will hold on the
defection strategies. However, the intermediate region will change
their strategies under the influence of their neighbors. Hence, the
cooperators' frequency of intermediate region is 0.5.
Figure~\ref{Fig4} confirms our conclusions discussed above, i.e.,
the isolation of communication among hubs in the structured
scale-free network decreases the emergence of cooperation.

\section{Conclusion}
In this paper, we have investigated the evolution of cooperation on
the structured scale-free network for the PD game. In contrast to
the conventional wisdom that the cooperation dominates on the BA
scale-free network, the cooperative behavior is inhibited on the
structured scale-free network. We find that neither the scale-free
property nor the high clustering coefficient is the determinant
factor for the inhibition of cooperation spreading on the structured
scale-free network. Then what is the root cause? Comparing with the
BA network, the age correlations are greatly suppressed on the
structured scale-free network. In result, hubs are usually linked to
low-degree nodes and simulations have confirmed the result that the
node-distance among hubs are much larger on the structured
scale-free network than on the BA network. Furthermore, to explain
heuristically why the spread of cooperation is inhibited, we
simplify the structured scale-free network as a double-star
connected by an intermediate region, which is formed by low-degree
nodes. Through detailed analysis we showed that the lack of age
correlations and associated `large-world' behavior are responsible
for the inhibition of cooperation on the structured scale-free
network.

\section*{Acknowledgment}

This research was supported by the National Basic Research Program
of China under grant No. 2007CB310806, the National Natural Science
Foundation of China under Grant Nos. 60704044, 60873040 and
60873070, Shanghai Leading Academic Discipline Project No. B114, and
the Program for New Century Excellent Talents in University of China
(NCET-06-0376). Rong Zhihai acknowledges support from the National
Science Foundation of P. R. China under Grant No. 60874089 and
70701009, the Shanghai Education Development Foundation under Grant
No. 2008CG38.

\end{document}